\documentclass[manuscript=article,journal=jpcafh]{achemso}
\usepackage[sort&compress,numbers,super]{natbib}
\usepackage{achemso}
\mciteErrorOnUnknownfalse

\usepackage[dvipsnames]{xcolor}
\usepackage{subcaption}
\usepackage{amssymb}
\usepackage{graphicx} 
\usepackage{subcaption} 
\usepackage{makecell}
\usepackage{amsmath}
\usepackage{braket}
\usepackage{gensymb}
\usepackage{multirow}
\usepackage{booktabs}
\usepackage{adjustbox}
\usepackage{array}
\usepackage{bm}
\usepackage{colortbl}
\usepackage{xcolor}
\usepackage{tabularx}
\usepackage{multirow}
\usepackage{multicol}
\usepackage{lipsum} 
\usepackage{amsmath,amsthm}
\usepackage{url}
\usepackage{hyperref}
\usepackage{cleveref}
\usepackage{microtype}
\usepackage{underscore}

\definecolor{myred}{rgb}{1, 0.698, 0.698}
\definecolor{myorange}{rgb}{1,0.851,0.698}
\definecolor{myyellow}{rgb}{1, 0.973, 0.773}
\newcolumntype{L}[1]{>{\raggedright\let\newline\\\arraybackslash\hspace{0pt}}m{#1}}
\newcolumntype{C}[1]{>{\centering\let\newline\\\arraybackslash\hspace{0pt}}m{#1}}
\newcolumntype{R}[1]{>{\raggedleft\let\newline\\\arraybackslash\hspace{0pt}}m{#1}}
\usepackage{pdfpages}
\usepackage[version=3]{mhchem}
\usepackage{xr}

\makeatletter
\newcommand*{\addFileDependency}[1]{
  \typeout{(#1)}
  \@addtofilelist{#1}
  \IfFileExists{#1}{}{\typeout{No file #1.}}
}
\makeatother

\newcommand*{\myexternaldocument}[1]{%
    \externaldocument{#1}%
    \addFileDependency{#1.tex}%
    \addFileDependency{#1.aux}%
}

\myexternaldocument{SI}

\title{Bridging Textual Descriptors and Molecular Geometries via Modality Transformation for Efficient Catalytic Configuration Identification}

\title{Distilling Geometric Knowledge into Textual Representations via Modality Transformation for Efficient Catalytic Configuration Identification}

\author{Zhenzhong Wang}
\affiliation{Xiamen University}

\author{Xiaobo Shang}
\affiliation{Xiamen University}

\author{Xiaojing Liu}
\affiliation{Xiamen University}

\author{Yiu-ming Cheung}
\affiliation{Hong Kong Baptist University}

\author{\newline Min Jiang}%
\affiliation{Xiamen University}
\email{minjiang@xmu.edu.cn}

\date{\today}

\begin{document}

\maketitle

\begin{abstract}
Discovering the optimal catalytic configuration with global minimal energy is central to catalyst design. Conventionally, this requires computationally expensive theoretical calculations across a large pool of candidates to identify the global minimum, creating a fundamental efficiency bottleneck. Recently, geometric deep learning has emerged as a promising alternative, yet its computational cost in processing complex atomic geometries remains an impediment to large-scale high-throughput screening. By contrast, language models operate directly on compact expert-crafted textual descriptors, offering faster inference but sacrificing prediction accuracy. In this work, we propose a cross-modal transformation framework that distills atomic geometry knowledge into compact textual representations. We formulate this distillation as a coarse-to-fine optimal transport problem: we first establish distribution-level correspondences between the two modalities and then refine instance-level relationships within them. Through this transformation, the framework operates directly on textual descriptors and bypasses expensive processing of atomic geometries during inference, improving efficiency while maintaining prediction accuracy. Experimental results demonstrate that the proposed model achieves a competitive mean absolute error (MAE) of 0.515 eV on the IS2RE task, outperforming both geometry-based and text-only baselines across most training scales. Furthermore, on the OC20Dense screening benchmark, it achieves a Top-1 hit rate of 15.52\%, with a median energy gap of only 0.10 eV relative to DFT-calculated energies, while maintaining an inference throughput of over 340 samples per second. These results highlight the model's ability to reconcile predictive accuracy with computational efficiency.
\end{abstract}

\flushbottom
\maketitle

\thispagestyle{empty}

\section{Introduction}

\begin{figure*}[t]
    \centering
    \includegraphics[width=1.0\textwidth]{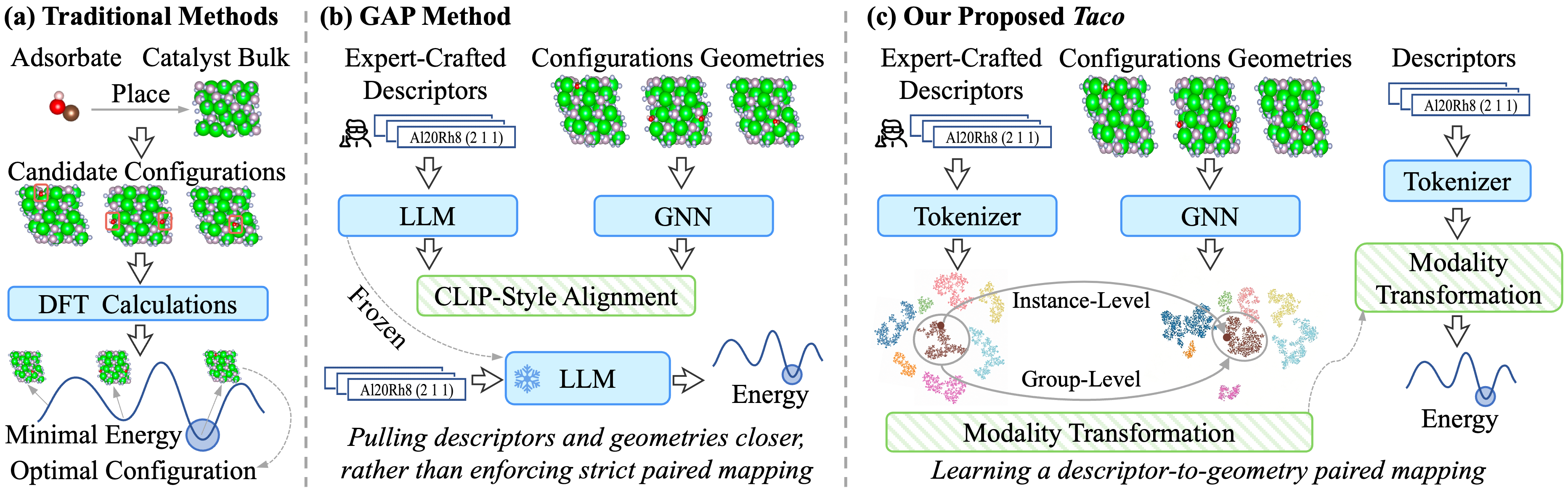}
    \caption{(a) Traditional methods rely on first-principles calculations such as DFT to compute the precise 3D structures and adsorption energies of a large pool of candidate configurations, from which the one with the global minimum energy is selected as the most stable. (b) GAP offers a pathway to bypass these costly geometric computations by pulling textual and geometric embeddings closer within a shared embedding space using a CLIP-style objective, but it does not provide a direct mapping from text to geometry. (c) Our {\em Taco} performs hierarchical transformation at both group and instance levels, establishing an explicit modality transformation from text to geometry that enhances both computational efficiency and predictive accuracy.}
    \vspace{-0.3cm}
    \label{fig:mappingIntro}
\end{figure*}

Accurately and efficiently identifying the optimal adsorption configuration corresponding to the global minimum in adsorption energy is a cornerstone of catalyst design~\cite{wang2024crystalline,pei2025non,bozal2025developing}. As shown in Figure \ref{fig:mappingIntro}(a), for a given adsorbate-catalyst system, a broad range of candidate adsorption configurations can be produced by enumerating possible adsorbate placement poses on the catalyst, typically based on expert knowledge or heuristic strategies. These candidates are distinguished by adsorption sites, molecular orientation, etc. Conventional identification of the optimal configuration largely depends on exhaustive density functional theory (DFT) calculations~\cite{argaman2000density}. Nevertheless, the formidable computational overhead severely limits its applicability to high-throughput screening workflows. This bottleneck has spurred the rapid advancement of data-driven surrogate models to accelerate computational catalyst exploration~\cite{yao2024property,wang2026explainablemolecularpropertyprediction,11200037,11123751,chanussot2021open,10494027,10595522}.

Machine learning has offered efficient alternatives for the discovery of molecules and materials~\cite{kolluru2024adsorbdiff,fung2021machine,ock2026adsorbAgent,liu2025machine}. These methods, which are often distinguished by how they represent the molecule system, can be roughly summarized into two lines. One line adopts geometric neural networks, which treat the system as an atomistic graph and learn from atomic positions, local neighborhoods, and geometric relationships~\cite{gilmer2017neural,xie2018crystal,joshi2025all,schutt2018schnet}. Among others, architectures such as DimeNet/DimeNet++\cite{gasteiger2020fast,klicpera2020dimenet}, and EquiformerV2~\cite{liao2024equiformerv2} further improved performance by explicitly modeling distances, angles, and symmetry‑aware interactions. 
The other line of work leverages language models by converting the molecular system into textual descriptors and performing prediction directly from the text~\cite{wang2026explainablemolecularpropertyprediction,zheng2025large,li2025large,li2025chemvlm,cavanagh2026smileyllama,wang2025token}. These methods demonstrate the potential to bypass 3D geometries using textual descriptors. For instance, CatBERTa shows that chemically meaningful text descriptors alone can support relaxed adsorption energy prediction~\cite{ock2023catberta}. Building on CatBERTa, as illustrated in Figure \ref{fig:mappingIntro}(b), GAP aligns geometric structures and textual descriptors via a CLIP‑style framework~\cite{ock2024multimodal,radford2021clip}. After alignment, predictions rely solely on textual descriptors without requiring explicit geometric encoding at inference time.


Despite these advances, both paradigms are constrained by an inherent accuracy–efficiency trade-off. Geometric deep learning models achieve high predictive accuracy by explicitly encoding 3D atomic coordinates and bond geometries, but their reliance on iterative message passing and aggregation over atomic graphs introduces substantial computational overhead~\cite{wang2025harnessing}. 
By contrast, language models operate directly on chemically structured textual descriptors, bypassing explicit 3D structure processing. This yields faster inference, yet the omission of explicit geometries typically results in lower accuracy and limited generalization to out-of-distribution adsorption systems~\cite{ock2023catberta}.

Improving efficiency and accuracy simultaneously is challenging, mainly because adsorption energy is highly sensitive to subtle atomic arrangements, whereas compact textual descriptors cannot capture such geometric differences. To balance the two, each textual descriptor—or its learned representation—should map precisely onto the geometry of its corresponding configuration.
This fundamentally distinguishes it from the GAP method depicted in Figure~\ref{fig:mappingIntro}(b). GAP aligns geometric and textual descriptors via a CLIP-style framework~\cite{ock2024multimodal}, which merely pulls paired text–geometry embeddings closer in a shared space. Such contrastive alignment encourages proximity between paired embeddings but does not specify how a textual representation should be transformed into its geometric counterpart.
Given that adsorption energy is highly sensitive to subtle structural details, this limitation leaves substantial room for improvement.

However, establishing this transformation poses a fundamental challenge: the structural inconsistency between the two representation spaces.
Although each textual descriptor is paired with a corresponding 3D configuration, independently trained text and geometric encoders organize these samples according to modality-specific characteristics. Consequently, their learned representations exhibit different distributions. This discrepancy makes cross-modal transformation more challenging than simply matching paired instances, calling for a mechanism that accounts for both the distributions and fine-grained correspondence of the two modalities.

To this end, we propose \emph{Taco}, a cross-modality framework that learns a correspondence between \underline{t}extual descriptors and their \underline{co}rresponding \underline{a}dsorption geometries. 
In particular, we formulate this text-geometry representation transformation as a knowledge distillation process, where the transfer of geometric information into the text domain is framed as a distribution matching problem, and establish this correspondence through optimal transport.
Specifically, we develop a Group-Conditioned Optimal Transport (GCOT) framework to address the mismatch between modality-specific representation structures. Rather than directly aligning paired instances, GCOT first characterizes the two modalities through soft groups and establishes their distributional correspondences at the group level. Conditioned on these group correspondences, it further performs instance-level transport to resolve fine-grained relationships between textual and geometric representations. By jointly modeling group-level structural correspondence and instance-level distributional matching, GCOT provides a hierarchical mechanism that effectively distills structural geometric knowledge into the textual representations, aligning them with the geometric latent space.
As presented in Figure \ref{fig:mappingIntro}(c), textual and geometric representations are first associated with a set of groups, producing group-wise distributions in both modalities. GCOT then performs group-level transport to identify structural correspondences between the two modalities, followed by instance-level distributional transport within each group pair to establish fine-grained cross-modal relationships.
This transformation mechanism effectively bridges the modality gap between textual and geometric representations, allowing accurate adsorption energy prediction using text inputs. 
\emph{Taco}  achieves competitive performance on the OC20 IS2RE task, reducing the mean absolute error to 0.515 eV under the full 460k training setting. Across most training scales, it outperforms both geometry-based and text-only baselines on the in-domain and out-of-distribution splits.
Moreover, on the OC20Dense high-throughput screening benchmark, Taco attains a Top-1 hit rate of 15.52\% and a median gap of 0.10 eV, demonstrating that the learned distilled text-to-geometry mapping enables efficient and accurate identification of the most stable catalytic configurations without requiring explicit geometric encoding at inference time.







\section{Results and Discussion}

\textbf{Problem definition.} 
For a given adsorbate-catalyst pair, i.e., with the same adsorbate and the same catalyst bulk, there exists a candidate set of adsorption configurations $\mathbb{C} = \{C_1, C_2, \dots, C_M\}$, where the configurations vary only in adsorption sites, molecular orientations, and local atomic arrangements. 
Each configuration $C_m$ admits a geometric graph $\mathcal{G}_m$ and a corresponding textual descriptor $\mathcal{S}_m$.
The adsorption energy prediction task aims to train a predictor $f_{\theta}$ to estimate the adsorption energy $\hat{E}_{\mathrm{ads},m}$ for each configuration, using $\mathcal{G}_m$, $\mathcal{S}_m$, or both as input. 
The screening task can be reduced to identifying the configuration with the lowest predicted energy: $C^* = \arg\min_{C_m \in \mathbb{C}} \hat{E}_{\mathrm{ads},m}$. 
The selected $C^*$ is taken as the most stable adsorption structure among the candidates for the given adsorbate-catalyst system.

\textbf{Data representations.} 
The adsorption configurations can be represented in two types of modalities: 3D geometries and textual descriptors. 
For the geometric representation, each adsorption system is modeled as an atomistic graph $\mathcal{G}=(\mathcal{V},\mathcal{E})$, where $\mathcal{V}=\{v_1,\dots,v_N\}$ denotes the atom set with coordinates $\mathbf{X}=[\mathbf{x}_1,\dots,\mathbf{x}_N]^T\in\mathbb{R}^{N\times 3}$. 
Under periodic boundary conditions, the edge set is constructed as $\mathcal{E}=\{(i,j,\mathbf{k}) \mid \|\mathbf{x}_i-(\mathbf{x}_j+\mathbf{L}\mathbf{k})\|_2 \le R_{\mathrm{cut}}\}$, where $\mathbf{L}$ is the lattice matrix, $\mathbf{k}$ is the cell offset vector, and $R_{\mathrm{cut}}$ is the cutoff radius. 
Among the candidate neighbors within $R_{\mathrm{cut}}$, the $K$ nearest neighbors are retained. In recent years, a substantial body of work has operated on such atomistic graphs, leveraging local coordination and geometric information to predict adsorption energies \cite{gilmer2017neural,xie2018cgcnn,schutt2018schnet,klicpera2020dimenet,gasteiger2020dimenetpp,gasteiger2021gemnet,gasteiger2022gemnetoc}. However, these approaches rely on iterative message passing and aggregation over atomic graphs, which introduces considerable computational overhead.

\begin{figure*}[t]
    \centering
    \includegraphics[width=1.0\textwidth]{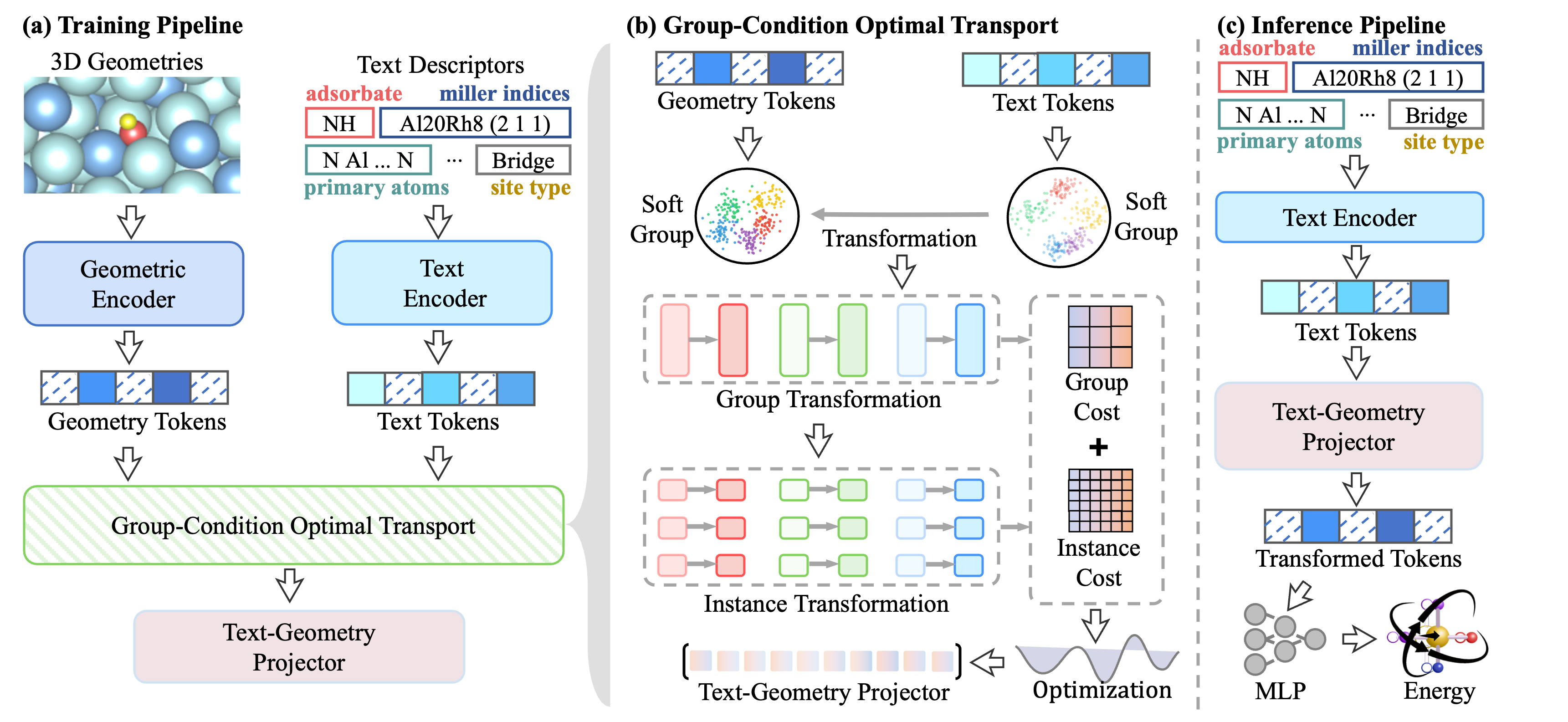}
    \caption{Overview of the proposed {\em Taco}. (a) At the training stage, the geometries and textual descriptors are input into a geometric encoder and a text encoder, producing geometry tokens and text tokens, respectively. These tokens are then transformed via a group-condition optimal transport module, yielding a modality transformation, i.e., a text-geometry projector. (b) The group-condition optimal transport module operates at the group level and the instance level, establishing cross-modal mapping between text descriptors and their corresponding geometries. (c) At the test stage, textual descriptors are fed into the text encoder. The learned projector then generates transformed tokens for energy prediction.}
    \vspace{-0.3cm}
    \label{fig:framework}
\end{figure*}

For the textual descriptors, as shown in Figure~\ref{fig:framework}(a), Ock \textit{et al.} designed descriptors that include the adsorbate formula $\texttt{Ad}$, surface elements $\texttt{Su}$, Miller indices $\texttt{Mi}$, primary atoms $\texttt{Pr}$, site types $\texttt{Si}$, and secondary atoms $\texttt{Se}$~\cite{ock2023catberta}. 
These attributes can be concatenated in a fixed order and tokenized as textual tokens $\mathcal{S}=\operatorname{Tokenizer}(\texttt{Ad} \oplus \texttt{Su} \oplus \texttt{Mi} \oplus \texttt{Pr} \oplus \texttt{Si} \oplus \texttt{Se})$. 
It is worth noting that the textual descriptor is deterministically generated from each 3D adsorption configuration following a predefined rule-based protocol. Specifically, the adsorbate symbol and catalyst composition are directly extracted from the geometric configuration, the Miller index specifies the crystallographic surface orientation, and the interacting atoms are identified according to covalent-radius criteria using the Pymatgen package~\cite{ong2013python}. Consequently, each adsorption configuration in the dataset is paired with a corresponding textual descriptor. 

Several studies have explored textual representations for prediction~\cite{wang2026explainablemolecularpropertyprediction,cao2023moformer,beltagy2019scibert,schwaller2019moleculartransformer,ghanekar2022acegcn,lee2020biobert}. Among them, CatBERTa~\cite{ock2023catberta} and GAP~\cite{ock2024multimodal} convert adsorption configurations into structured text sequences and show that textual descriptors alone can enable energy prediction.
However, this efficiency typically sacrifices accuracy, as these methods rely solely on textual descriptors and discard precise 3D geometries. GAP, for example, attempts to mitigate this by aligning paired text–geometry embeddings in a shared space. Yet such contrastive alignment only encourages proximity between the paired embeddings; it does not specify how a textual representation should be transformed into its geometric counterpart. Consequently, its predictive performance still leaves considerable room for improvement.




\subsection{Framework}

We therefore introduce Taco, a cross-modal framework that distills geometric knowledge into the text domain during training, aiming to enable accurate adsorption energy prediction directly from textual descriptors at inference time.
As shown in Figure~\ref{fig:framework}(a), the 3D geometries and textual descriptors are fed into a geometric neural network and a text encoder, producing geometry tokens and text tokens, respectively. Here, the geometric encoder serves as a teacher providing rich structural information, while the text encoder acts as a student learning to represent it.
These tokens are subsequently processed by GCOT, which acts as the distillation mechanism to explicitly account for the structural mismatch between the two modalities. Specifically, soft group assignments characterize the modality-specific organization of textual and geometric representations. GCOT then establishes cross-modal correspondences at two complementary levels: group-level transport identifies the distributional relationships between modality-specific groups, while instance-level transport resolves fine-grained correspondences conditioned on each group pair. 
Importantly, these two levels address different aspects of cross-modal transformation: the former captures the mismatch in representation organization, whereas the latter establishes fine-grained distributional correspondences between individual representations conditioned on each group pair.
By minimizing the optimal transport costs at both levels, we distill the geometric knowledge to learn a dedicated projector that maps textual representations into the geometric feature space. This learned projector serves as the core component that enables efficient and accurate inference using only textual descriptors.

At test time, as illustrated in Figure~\ref{fig:framework}(c), textual descriptors of candidate configurations are fed into the text encoder.
Subsequently, the learned text-geometry projector generates the corresponding transformed geometric tokens, based on which the energy head multilayer perceptron (MLP) performs energy prediction. 
By using the compact textual descriptors, {\em Taco} allows for rapid screening of adsorption configurations during inference. Detailed descriptions of the major components are provided in the Methodology section.


\subsection{Benchmark Datasets}

\textbf{OC20 IS2RE.} The Open Catalyst (OC) datasets mark a pivotal advancement in computational catalysis, comprising more than 1.2 million DFT structural relaxation samples that span 55 elemental species and 82 adsorbate types~\cite{chanussot2021open}, covering abundant nitrogen, carbon, and oxygen reaction chemistries over varied inorganic material surfaces.

In this work, we focus on the Initial Structure to Relaxed Energy (IS2RE) task in OC20. Given an unrelaxed initial adsorption structure, the goal is to predict the final relaxed adsorption energy without carrying out DFT relaxation. We evaluate the model under three training settings: Train-10k, Train-100k, and the full-scale 460k setting. These settings allow us to examine model behavior in both low-data and large-scale regimes. 
Since the DFT relaxation trajectories for the test dataset are not publicly accessible, our evaluation and comparison of model performance are centered on the validation dataset.
The database provides four official validation splits to systematically evaluate model generalization: The \text{ID} split contains adsorbates and catalysts seen during training but uses new geometries. The \text{OOD}$_{\text{ads}}$ split contains unseen adsorbates on known catalysts, while the \text{OOD}$_{\text{cat}}$ split contains known adsorbates on unseen catalyst compositions. The \text{OOD}$_{\text{both}}$ split contains both unseen adsorbates and unseen catalyst materials.


\textbf{MamunHighT2019.} To complement the OC datasets, we evaluate on the CatBench-formatted MamunHighT2019 adsorption dataset~\cite{winther2019catalysis}, which is curated from Catalysis-Hub and provides DFT-referenced adsorption energies for heterogeneous catalysis. Each configuration contains an adsorbate–surface structure and a reference adsorption energy. The processed dataset contains 45,130 unique configurations, spanning 15 adsorbates across 1,916 bulk materials. We construct a reproducible train and validation protocol with one training split and four validation subsets, namely \text{ID}, \text{OOD}$_{\text{ads}}$, \text{OOD}$_{\text{cat}}$, and \text{OOD}$_{\text{both}}$, defined by adsorbate and catalyst-composition overlap with the training set.


\textbf{OC20Dense.} Beyond energy prediction on OC20 IS2RE and MamunHighT2019, we also evaluate high-throughput adsorption configuration screening on OC20Dense. Unlike standard IS2RE, where each adsorption structure is treated as an independent regression example, OC20Dense provides multiple candidate adsorption configurations for the same adsorbate-catalyst system. 
Candidate configurations are ranked by predicted relaxed adsorption energy, and the lowest-energy configuration selected by the model is compared with the DFT global minimum. This setting tests whether a model can identify the most stable adsorption configuration from a dense candidate pool, a central requirement in practical catalyst screening. 

For detailed information about all the datasets, please refer to the Supplemental Materials.

\begin{table*}[htbp]
  \centering
\caption{MAE for energy prediction across various models on the OC20 IS2RE dataset. The best result is \textbf{bold}.}
  \resizebox{0.7\textwidth}{!}{
    \small
    \begin{tabular}{llcccc}
    \toprule

    \multicolumn{1}{c}{\text{Data Size}} & \multicolumn{1}{c}{\text{Model}} & \text{ID} & \text{OOD}$_{\text{ads}}$ & \text{OOD}$_{\text{cat}}$ & \text{OOD}$_{\text{both}}$ \\
    \midrule
    10k   & Graphormer~\cite{ying2021transformers}   & 0.962±0.003 & 0.972±0.002 & 0.977±0.011 & 0.990±0.003 \\
    10k   & CatBERTa~\cite{ock2023catberta}  & 0.770±0.011 & 0.996±0.003 & \textbf{0.713±0.013} & 0.933±0.011 \\
    10k   & GAP~\cite{ock2024multimodal}  & 0.830±0.009 & 0.841±0.004 & 0.843±0.007 & 0.839±0.013 \\
    10k   & CGCNN~\cite{xie2018crystal}  & 1.045±0.026 & 0.975±0.014 & 1.028±0.005 & 0.862±0.004 \\
    10k   & DimeNet++~\cite{klicpera2020dimenet} & 0.949±0.016 & 0.890±0.014 & 0.888±0.007 & 0.808±0.004 \\
    10k   & EquiformerV2~\cite{liao2024equiformerv2}  & 0.783±0.013 & 0.797±0.017 & 0.734±0.009 & 0.776±0.006 \\
    \rowcolor[rgb]{ .851,  .851,  .851} 10k   & \textit{\textbf{Taco}} & \textbf{0.770±0.023} & \textbf{0.764±0.050} & 0.740±0.013 & \textbf{0.749±0.015} \\
    \midrule
    100k  & Graphormer~\cite{ying2021transformers}   & 0.638±0.007 & 0.793±0.012 & 0.650±0.006 & 0.749±0.011 \\
    100k  & CatBERTa~\cite{ock2023catberta} & 0.678±0.012 & 0.944±0.006 & 0.647±0.016 & 0.897±0.008 \\
    100k  & GAP~\cite{ock2024multimodal} & 0.699±0.004 & 0.893±0.008 & 0.665±0.014 & 0.856±0.012 \\
    100k  & CGCNN~\cite{xie2018crystal} & 0.988±0.017 & 0.824±0.011 & 0.925±0.003 & 0.759±0.002 \\
    100k  & DimeNet++~\cite{klicpera2020dimenet}  & 0.812±0.013 & 0.788±0.011 & 0.759±0.006 & 0.718±0.005 \\
    100k  & EquiformerV2~\cite{liao2024equiformerv2} & 0.670±0.009 & \textbf{0.781±0.011} & 0.614±0.008 & 0.703±0.003 \\
    \rowcolor[rgb]{ .851,  .851,  .851} 100k  & \textit{\textbf{Taco}} & \textbf{0.652±0.019} & 0.795±0.024 & \textbf{0.604±0.015} & \textbf{0.680±0.017} \\
    \midrule
    460k  & Graphormer~\cite{ying2021transformers}  & 0.612±0.007 & 0.754±0.008 & 0.614±0.003 & 0.720±0.006 \\
    460k  & CatBERTa~\cite{ock2023catberta} & 0.643±0.004 & 0.905±0.013 & 0.601±0.008 & 0.869±0.008 \\
    460k  & GAP~\cite{ock2024multimodal} & 0.671±0.005 & 0.850±0.004 & 0.626±0.006 & 0.838±0.002 \\
    460k  & CGCNN~\cite{xie2018crystal}  & 0.944±0.011 & 0.783±0.008 & 0.861±0.006 & 0.689±0.004 \\
    460k  & DimeNet++~\cite{klicpera2020dimenet} & 0.747±0.009 & 0.738±0.006 & 0.710±0.012 & 0.685±0.009 \\
    460k  & EquiformerV2~\cite{liao2024equiformerv2} & 0.589±0.009 & 0.599±0.002 & 0.552±0.005 & 0.562±0.004 \\
    \rowcolor[rgb]{ .851,  .851,  .851} 460k  & \textit{\textbf{Taco}} & \textbf{0.515±0.016} & \textbf{0.577±0.008} & \textbf{0.539±0.009} &\textbf{ 0.557±0.014} \\
    \bottomrule
    \end{tabular}%
    }
  \label{tab:is2re}%
\end{table*}%

\begin{table*}[t]
\centering
\caption{MAE of different models on the MamunHighT2019 dataset. The best result is shown in \textbf{bold}.}
\setlength{\tabcolsep}{3pt}
 \resizebox{0.57\textwidth}{!}{
    \small
\begin{tabular}{l cccc}
\toprule
Model
& ID & OOD$_{\text{ads}}$ & OOD$_{\text{cat}}$ & OOD$_{\text{both}}$ \\
\midrule

Graphormer~\cite{ying2021transformers} 
& 0.142${\pm0.014}$ & 0.163${\pm0.006}$ & 0.155${\pm0.012}$ & 0.172${\pm0.003}$ \\

CatBERTa~\cite{ock2023catberta} 
& 0.195${\pm0.004}$ & 0.218${\pm0.010}$ & 0.207${\pm0.008}$ & 0.228${\pm0.013}$ \\

GAP~\cite{ock2024multimodal} 
& 0.159${\pm0.011}$ & 0.175${\pm0.007}$ & 0.167${\pm0.014}$ & 0.183${\pm0.008}$ \\

CGCNN~\cite{xie2018crystal} 
& 0.155${\pm0.007}$ & 0.174${\pm0.013}$ & 0.164${\pm0.005}$ & 0.182${\pm0.010}$ \\

DimeNet++~\cite{klicpera2020dimenet} 
& 0.138${\pm0.008}$ & 0.156${\pm0.011}$ & 0.147${\pm0.006}$ & 0.167${\pm0.015}$ \\

EquiformerV2~\cite{liao2024equiformerv2} 
& 0.131${\pm0.012}$ & 0.149${\pm0.009}$ & 0.140${\pm0.015}$ & 0.160${\pm0.004}$ \\

\rowcolor[gray]{0.9}
\textit{\textbf{Taco}}
& \textbf{0.124±0.012} & \textbf{0.140±0.017} & \textbf{0.135±0.014} & \textbf{0.153±0.013} \\
\bottomrule
\end{tabular}
}
\label{tab:mamun}
\end{table*}



\subsection{Baselines} 

We compare \emph{Taco} with three categories of baselines: (1) pure text-based method CatBERTa~\cite{ock2023catberta}; (2) graph-based methods, including CGCNN~\cite{xie2018crystal}, DimeNet++~\cite{klicpera2020dimenet}, EquiformerV2~\cite{liao2024equiformerv2}, and Graphormer~\cite{ying2021transformers}; and (3) a multimodal baseline, GAP~\cite{ock2024multimodal}. Among these, CGCNN and DimeNet++ utilize officially available IS2RE results or pre-trained checkpoints. For Graphormer, CatBERTa, GAP, and EquiformerV2, since pre-trained weights for the IS2RE task have not been released, we trained them using publicly available source code. Detailed hyperparameter settings are provided in the Supplemental Materials.

\subsection{Text Descriptors-Based IS2RE Prediction}

\textbf{Prediction accuracy.} As summarized in Table~\ref{tab:is2re}, \emph{Taco} demonstrates competitive performance across diverse evaluation scenarios. When trained on the low-data regime of 10k samples, \emph{Taco} achieves the lowest overall MAE and outperforms all baseline models on the in-domain, OOD$_{\text{ads}}$, and OOD$_{\text{both}}$ subsets, while remaining highly competitive on OOD$_{\text{cat}}$. Under the 100k configuration, our method continues to secure the best results on OOD$_{\text{cat}}$ and OOD$_{\text{both}}$ subsets, while remaining highly competitive on the in-domain split. For the full 460k training set, \emph{Taco} achieves lower MAE values on the in-domain, OOD$_{\text{ads}}$, OOD$_{\text{cat}}$, and OOD$_{\text{both}}$.



\begin{figure*}[!t]
    \centering
    \vspace*{-0.3cm}
    \includegraphics[width=1.0\textwidth]{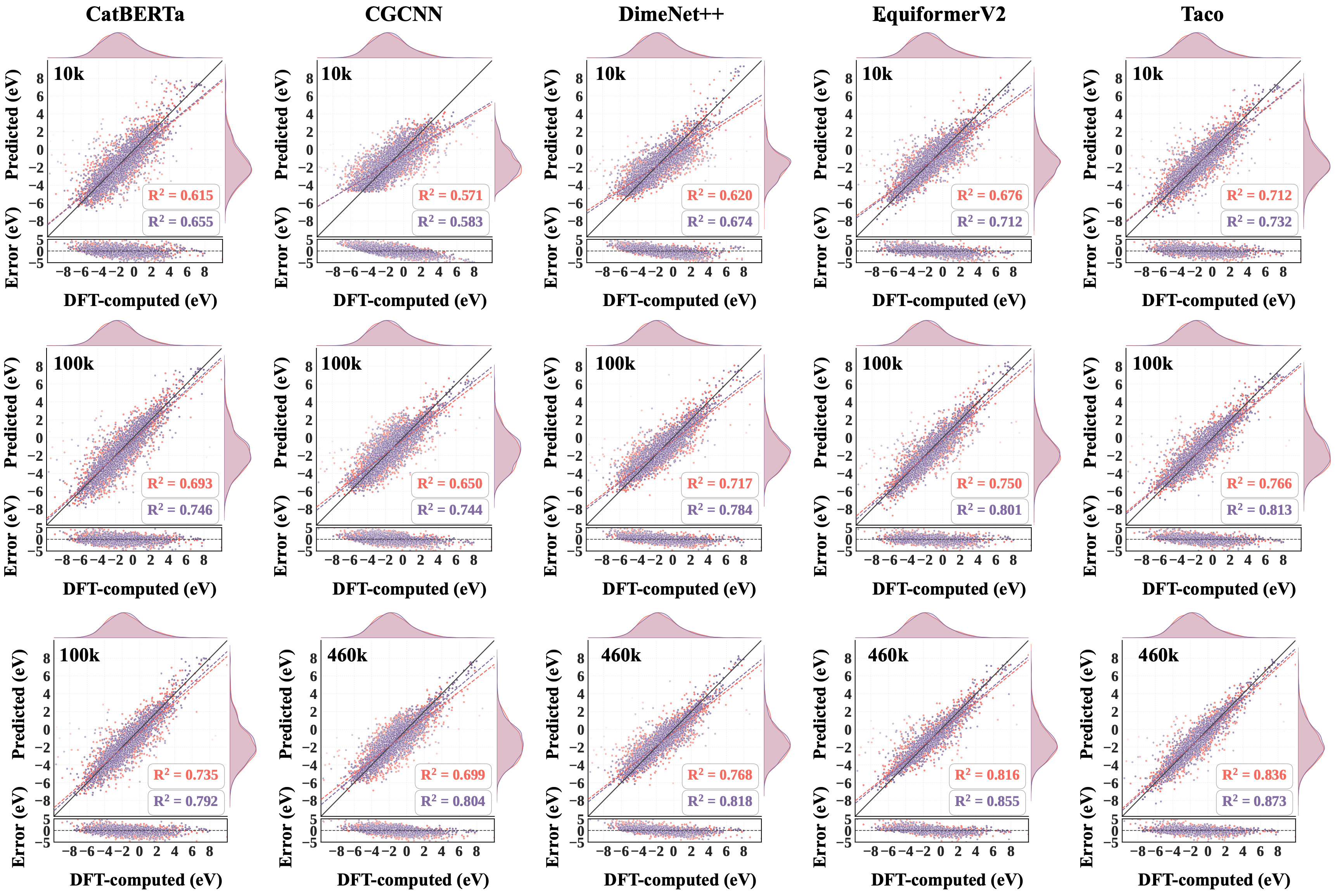}
    \vspace{-0.5cm}
    \caption{Coefficient of determination $R^2$. Pink and purple dots denote in-domain and OOD$_{\text{cat}}$ data, respectively. 
    }
    \vspace{-0.1cm}
    \label{tacoR2460}
\end{figure*}

To further assess the transferability of our method, we conduct additional validation on the MamunHighT2019 dataset. As reported in Table~\ref{tab:mamun}, \emph{Taco} consistently achieves the lowest MAE across all evaluation splits, including ID, OOD$_{\text{ads}}$, OOD$_{\text{cat}}$, and OOD$_{\text{both}}$, with values of 0.124, 0.140, 0.135, and 0.153 respectively. Our method surpasses the strongest baseline EquiformerV2 by relative improvements of 5.3\% on the challenging OOD$_{\text{both}}$ subset, demonstrating robust generalization beyond the OC20 training distribution.

We further adopt the coefficient of determination $R^2$ to quantify the correlation between predicted adsorption energies and reference values obtained from DFT calculations. Scatter diagrams corresponding to models trained with 10k, 100k, and 460k samples are illustrated in Figure~\ref{tacoR2460}, where pink points denote in-domain samples and purple points represent data from the $\text{OOD}_\text{cat}$ category.
The proposed \emph{Taco} maintains superior $R^2$ performance on both in-domain and out-of-distribution catalyst samples under all training volumes. When trained on 10k samples, our method achieves an in-domain $R^2$ of 0.712 and an $\text{OOD}_\text{cat}$ $R^2$ of 0.732, outperforming geometric-based EquiformerV2, as well as text-only CatBERTa with scores of 0.615 and 0.655. With full-scale 460k training data, our approach attains in-domain and out-of-distribution $R^2$ values of 0.836 and 0.873 respectively. Such results exceed those yielded by EquiformerV2, DimeNet++, CGCNN and CatBERTa.


Experimental results show that \emph{Taco} achieves competitive accuracy compared with purely geometric networks. 
Pure geometric models learn representations directly from 3D atomic structures, but the geometric configurations contain a large amount of irrelevant information, e.g., features from the bulk. In contrast, compact textual descriptors explicitly encode some key physicochemical factors that govern adsorption behavior, and using the text descriptors could directly capture the most relevant features.
Through the learned transformation, the model combines the textual priors with the discriminative power of geometric representations, which together lead to promising predictive performance in some cases.


\subsection{Text Descriptors-based Configuration Screening}

\begin{table*}[htbp]
  \centering
  \caption{Adsorption configuration screening results on the OC20Dense benchmark. The best result is shown in \textbf{bold}.}
  \resizebox{0.82\textwidth}{!}{
    \small
    \begin{tabular}{lccccc}  
    \toprule
    \multicolumn{1}{c}{\text{Model}} & \text{Top-1 Hit \%} & \text{Top-2 Hit \%} & \text{Top-3 Hit \%} & \text{Median Gap} & \text{Forward Samples/s} \\
    \midrule
     Graphormer~\cite{ying2021transformers} & 11.96±1.59 & 20.51±1.21 & 31.84±1.45 & 0.10±0.006 & 209.17 \\
     
    CatBERTa~\cite{ock2023catberta} & 13.23±1.85 & 24.74±1.74 & 36.47±1.72 & 0.21±0.014 & \textbf{436.92} \\

    GAP~\cite{ock2024multimodal} & 13.79±2.57 & 26.90±1.39 & 37.79±1.70 & 0.33±0.025 & 404.80 \\
   
    CGCNN~\cite{xie2018crystal} & 12.51±1.53 & 23.22±1.42 & 33.18±1.58 & 0.18±0.033 & 326.05 \\
    DimeNet++~\cite{klicpera2020dimenet}   & 13.28±1.74 & 25.59±1.36 & 36.16±1.59 & 0.12±0.007 & 160.37 \\
    EquiformerV2~\cite{liao2024equiformerv2}  & 13.80±2.07 & 26.20±1.51 & 37.03±1.75 & 0.19±0.020 & 84.97 \\
    \rowcolor[rgb]{ .851,  .851,  .851} \textit{\textbf{Taco}} & \textbf{15.52±1.70} & \textbf{28.90±1.57} & \textbf{39.65±1.93} & \textbf{0.10±0.014} & 340.82 \\
    \bottomrule
    \end{tabular}%
    }
  \label{tab:screening}%
\end{table*}

\textbf{Screening capability.}
To assess the practical utility of \emph{Taco} in high-throughput screening, we evaluate its ability to identify the most stable adsorption configuration on the OC20Dense benchmark. All comparison models are trained on OC20 IS2RE tasks and applied directly without any task-specific retraining, providing a rigorous test of generalization under realistic conditions. 
We report the following three metrics:

\begin{enumerate}
    \item Top-$k$ hit rate: the fraction of systems for which the true global minimum is ranked among the model’s top‑$k$ most stable predictions ($k=1,2,3$).

    \item Median Gap: the median of the energy differences $\Delta E$ across all systems, where each $\Delta E$ is the DFT energy difference of the model-selected candidate relative to the true global minimum. 
    \item Inference throughput (Forward samples/s): the number of configurations processed per second, quantifying computational efficiency.
\end{enumerate}

\begin{figure*}[t!]
    \centering 
    \vspace*{-0.3cm}
     \hspace*{-0.6cm} 
    \includegraphics[width=1.0\textwidth]{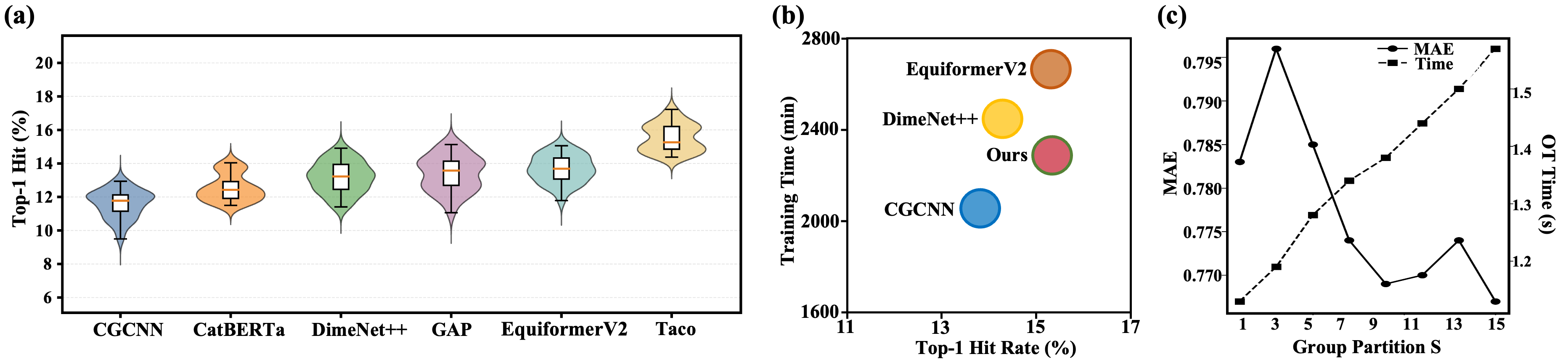}
   \caption{(a) Distribution of Top-1 hit rates across different algorithms in high-throughput computational screening experiments. (b) Top-1 hit rate and training efficiency on the 460k setting using different graph encoders. (c) Relationship between soft group number, MAE, and OT time cost on the OC20 IS2RE dataset of the 10K setting.}
    \label{tacoAblation}
\end{figure*}

Quantitative screening results are summarized in Table~\ref{tab:screening}. Our \emph{Taco} framework achieves the best performance across all ranking metrics, with Top-1, Top-2, and Top-3 hit rates of 15.52\%, 28.90\%, and 39.65\%, respectively, outperforming all baseline methods including the geometry-based EquiformerV2, the text-only CatBERTa, and the multimodal GAP. In terms of selection quality, \emph{Taco} achieves a Median Gap of 0.10 eV, tying with the best geometry-based baseline (Graphormer), indicating that our method identifies candidates with minimal energy penalty relative to the true global minimum. 
The statistical distribution of the top-1 hit rate is visualized in Figure~\ref{tacoAblation}(a). Our method obtains the highest median value, which outpaces DimeNet++, EquiformerV2, CatBERTa, CGCNN, and GAP, further confirming the robustness and superiority of our approach.

\begin{figure*}[t!]
    \centering 
    \includegraphics[width=1.0\textwidth]{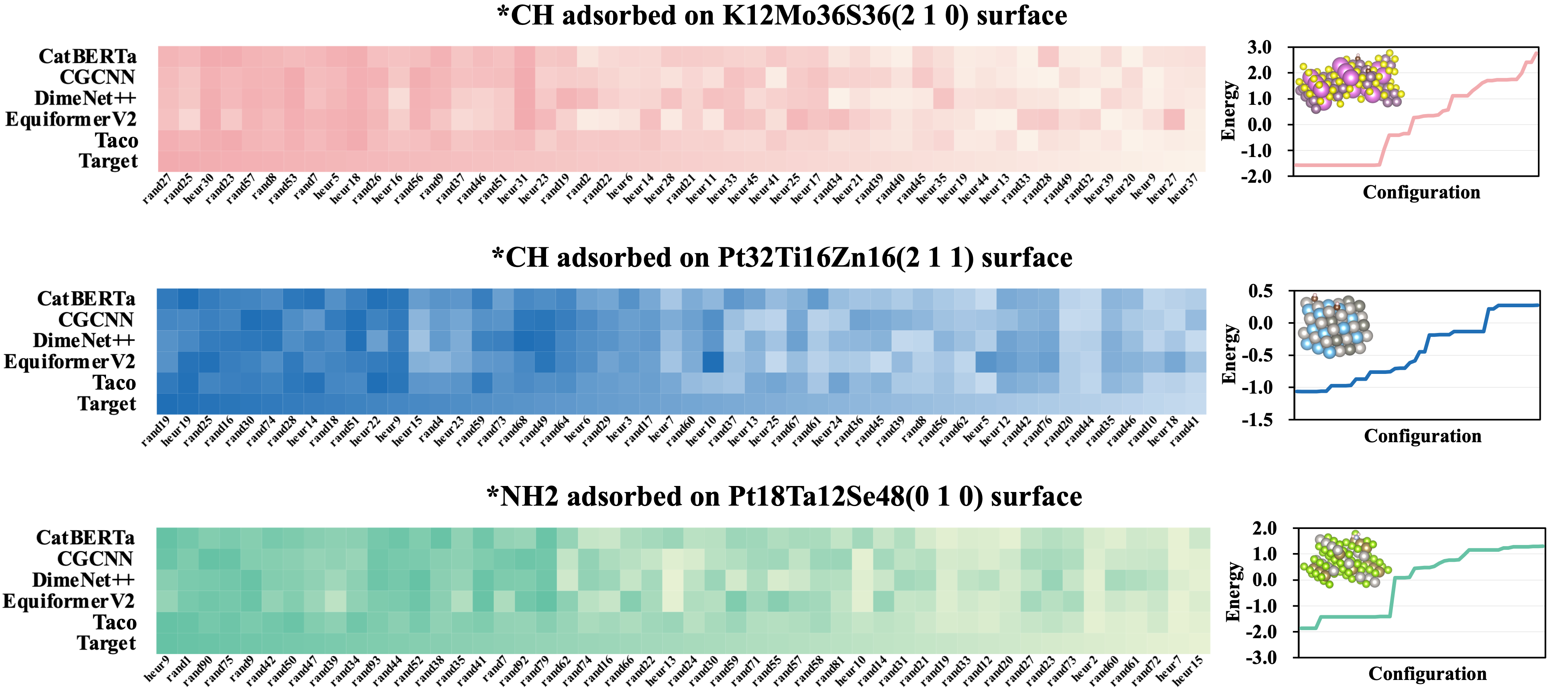}
    \vspace{-0.5cm}
    \caption{High-throughput screening validation on three adsorbate–catalyst systems. Left: Systems evaluated: (1) \ce{$\ast$CH} on \ce{K12Mo36S36(2 1 0)}; (2) \ce{$\ast$CH} on \ce{Pt32Ti16Zn16(2 1 1)}; (3) \ce{$\ast$NH2} on \ce{Pt18Ta12Se48(0 1 0)}. For each, over 40 initial configurations (heuristic and random) are evaluated with DFT-relaxed energies. Right: Corresponding energy landscapes exhibit three distinct patterns: steep transition, gradual slope, and bimodal distribution, providing a rigorous testbed for model robustness.}
    \vspace{-0.3cm}
    \label{screeningmat}
\end{figure*}

Beyond screening accuracy, computational efficiency is another key consideration for practical large-scale catalyst screening. All methods are tested on the same hardware and settings. Reported inference times include tokenization time and descriptor generation cost (The descriptor generation from a given 3D configuration requires only about 0.0015 seconds). As reported in Table~\ref{tab:screening}, \emph{Taco} achieves a throughput of 340.82 samples per second, notably higher than Graphormer, DimeNet++, and EquiformerV2, while remaining competitive with CGCNN and moderately behind CatBERTa and GAP. This efficiency advantage is particularly significant given that EquiformerV2, despite its strong predictive performance, suffers from lower inference speed due to its computationally intensive geometric processing. These results demonstrate that \emph{Taco} strikes a favorable balance between prediction accuracy and inference speed, rendering it well-suited for large-scale adsorption configuration screening where both effectiveness and computational cost matter.

\textbf{Ranking visualization.} To further validate the high-throughput screening capability, we evaluate its performance on three distinct adsorbate–catalyst systems with 40 initial configurations. Figure~\ref{screeningmat} presents the results for: (1) \ce{$\ast$CH} adsorbed on the \ce{K12Mo36S36(2 1 0)} surface; (2) \ce{$\ast$CH} adsorbed on the \ce{Pt32Ti16Zn16(2 1 1)} surface; and (3) \ce{$\ast$NH2} adsorbed on the \ce{Pt18Ta12Se48(0 1 0)} surface. 
The right-hand panels illustrate the energy landscape of each system, revealing diverse distributions: a steep energy transition, a gradual slope, and a gradually rising plateau with intermittent sharp jumps, respectively. These varied landscapes provide a rigorous testbed for assessing model robustness across different structural and energetic regimes.

\begin{figure*}[t]
    \centering    \includegraphics[width=1.0\textwidth]{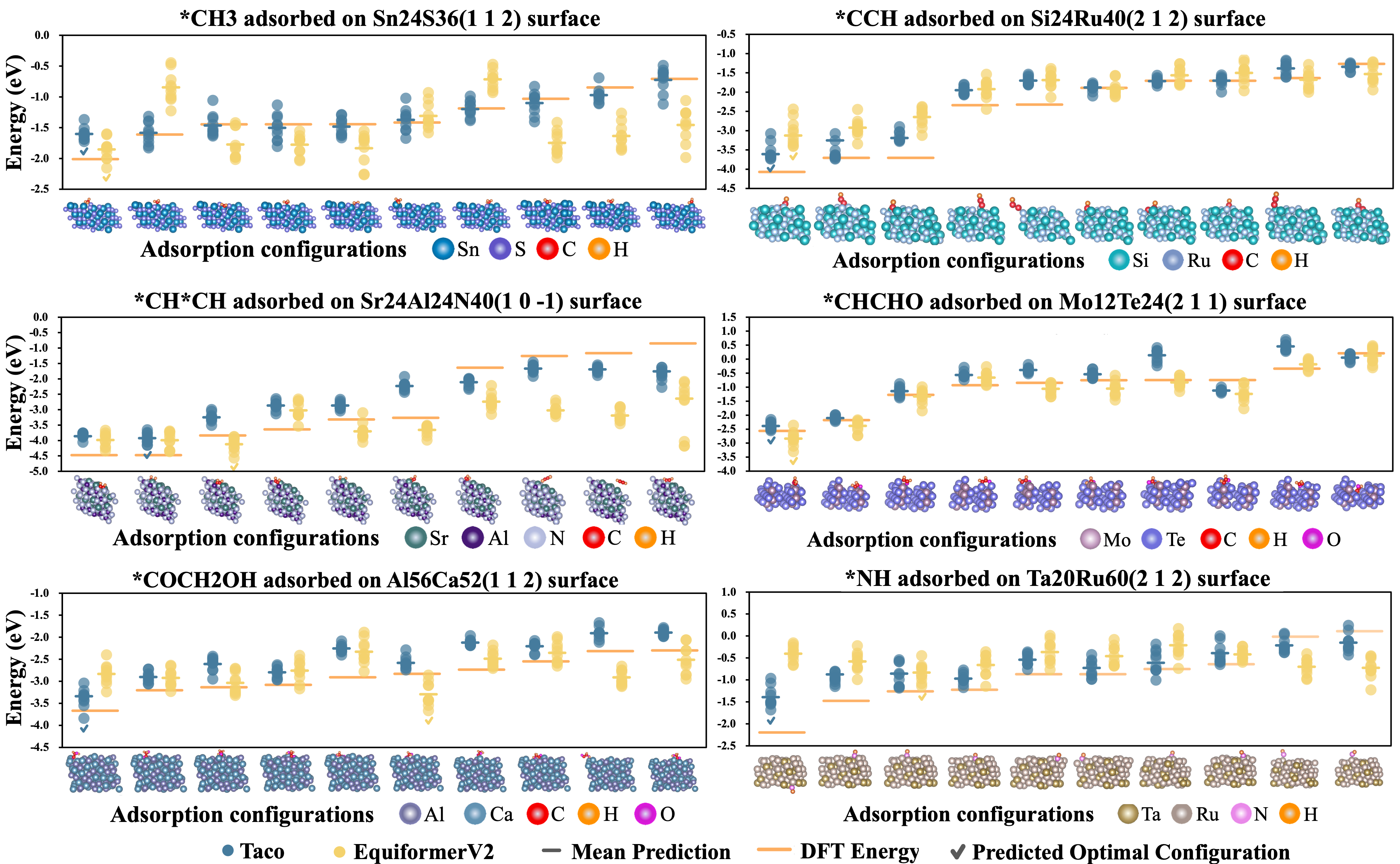}
    \vspace{-0.5cm}
    \caption{Predicted adsorption energies for six representative adsorbate-substrate systems. Each panel corresponds to a distinct adsorbate (\ce{$\ast$CH3}, \ce{$\ast$CCH}, \ce{$\ast$CH$\ast$CH}, \ce{$\ast$CHCHO}, \ce{$\ast$COCH2OH}, \ce{$\ast$NH}) on a different alloy surface. For each system, we compare the energy predictions of \emph{Taco} (blue) and EquiformerV2 (yellow) across multiple candidate configurations, with the DFT-calculated ground-truth energies shown as orange lines. }
\label{modelstability}
    \vspace{-0.3cm}
\end{figure*}

The heatmaps compare the predicted rankings from competing methods against the DFT ground truth, where darker colors correspond to lower predicted/ground truth energies and higher predicted/ground truth rankings. Across all three systems and across both low- and high-energy regimes, \emph{Taco} reproduces the ground-truth color pattern most faithfully. In particular, our model accurately identifies the dark-colored low-energy configurations, whereas baseline methods often fail to capture these key low-ranking structures, frequently miscoloring them as high-energy (light-colored) states. In contrast, \emph{Taco} not only matches the absolute color intensity but also preserves the relative ordering of configurations, correctly assigning the darkest hues to the most stable structures. These results demonstrate that \emph{Taco} reliably identifies promising candidates across diverse energy landscapes for adsorption configuration discovery.

\textbf{Screening stability.} To evaluate the prediction stability of \emph{Taco}, we compared it with the state-of-the-art EquiformerV2 across six chemically diverse adsorbate--catalyst pairs (Figure~\ref{modelstability}). These six systems were selected to span a broad range of adsorbate sizes (from single-carbon species such as \ce{$\ast$CH3} to multi-atom adsorbates like \ce{$\ast$COCH2OH}), and a variety of alloy surface compositions. This selection ensures a representative coverage of the chemical complexity encountered in typical heterogeneous catalysis applications. Each system features a distinct adsorbate (\ce{$\ast$CH3}, \ce{$\ast$CCH}, \ce{$\ast$CH$\ast$CH}, \ce{$\ast$CHCHO}, \ce{$\ast$COCH2OH}, \ce{$\ast$NH}) on a different alloy surface, with 10 distinct initial adsorption configurations per system. Both models were independently trained ten times using different random seeds, yielding ten sets of relaxed energy predictions for each configuration.

As shown, \emph{Taco} exhibits narrower prediction distributions across all six systems, in sharp contrast to the larger variances observed in EquiformerV2. This demonstrates improved stability and prediction consistency. Meanwhile, the mean prediction values of \emph{Taco} align closely with DFT-derived ground-truth energies, while EquiformerV2 shows noticeable deviations in some systems.
These results confirm that \emph{Taco} achieves both high prediction accuracy and stability.

\textbf{Ablation and sensitivity studies.} In this part, we conduct ablation studies on the IS2RE task to quantify the individual contributions of \emph{Taco}.
We first examine the adaptability of our framework by substituting the designed geometric encoder with other mainstream geometric encoders, including DimeNet++, EquiformerV2, and CGCNN, while keeping all other module configurations fixed. Figure~\ref{tacoAblation}(b) presents the Top-1 Hit and training cost. Here, we focus on comparing the training cost, as the geometric encoders are not used during inference, meaning their inference costs are consistent across all variants.
Among these geometric encoders, the proposed geometric encoder achieves a good trade-off between training time and Top-1 Hit rate, justifying the rationality of using the designed geometric encoder.

We then conduct ablation experiments to validate the complementarity of the two modalities and the effectiveness of our proposed transformation strategies. Two simplified variants are constructed for comparison: one purely relies on the designed geometric encoder to conduct energy regression without textual input, and the other removes geometric encoding branches and makes predictions merely based on chemical descriptions.
As shown in Table~\ref{tab:ablation_speed_accuracy}, the complete \emph{Taco} framework achieves the minimal MAE (0.515 eV), outperforming both single-modality variants. The geometry-only approach acts as a strong teacher, attaining competitive accuracy (0.544 eV) that already surpasses the SOTA geometric baseline EquiformerV2 (0.589 eV). However, it suffers from lower inference throughput due to the heavy geometric message passing. Conversely, the text-only model yields the worst accuracy (0.650 eV), indicating that text information alone cannot capture subtle structural details. Crucially, the 0.135 eV gap between Text-only (0.650 eV) and \emph{Taco} (0.515 eV) demonstrates that our GCOT framework successfully transfers the structural knowledge from the geometric teacher, rather than the text branch passively inheriting the teacher's performance.

Interestingly, \emph{Taco} even slightly outperforms its geometric teacher. This is because pure geometric encoders must suppress bulk-dominated irrelevant information from data, whereas compact textual descriptors explicitly encode key physicochemical factors (e.g., Miller index, site type) that govern adsorption. \emph{Taco} effectively fuses the distilled geometric knowledge with these strong textual priors, achieving a synergistic improvement. Ultimately, this distillation enables \emph{Taco} to inherit the teacher's capability while offering a promising direction to reduce reliance on computationally expensive 3D structure processing during inference.

\begin{figure}[t!]
    \centering 
  
     \hspace*{-0.6cm} 
\includegraphics[width=0.99\textwidth]{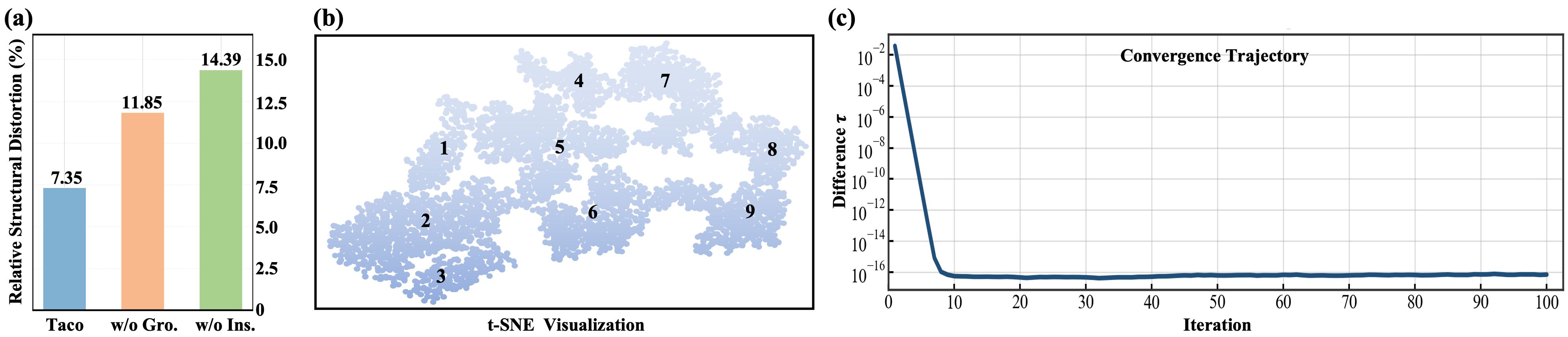}
   \caption{(a) Structural distortion error for the Taco and ablation variants. (b) Feature t-SNE visualization revealing 9 rough groups. (c) Convergence trajectory of the proposed optimizer.}
    \label{tacoTSNE}
     \vspace{-0.4cm}
\end{figure}

\begin{table}[t]
\centering
\caption{Prediction accuracy and inference speed of \emph{Taco} and its ablated variants on the IS2RE task.}
\label{tab:ablation_speed_accuracy}
\resizebox{0.463\textwidth}{!}{
\begin{tabular}{lcc}
\toprule
\text{Model / Variant} & \text{MAE (eV)} & \text{Forward Samples/s} \\
\midrule
Geometry-only            & 0.544 & 293.67 \\
Text-only                & 0.650 & 338.11 \\
w/o Group-level OT       & 0.703 & 335.86 \\
w/o Instance-level OT    & 0.692 & 342.16 \\
CLIP Alignment    & 0.677  &  330.47 \\
\rowcolor[gray]{0.9} \textit{\textbf{Taco}}       & 0.515 & 340.82 \\
\bottomrule
\end{tabular}
}
\end{table}

We further validate the efficacy of our proposed optimal transport transformation. If the text branch merely inherited knowledge without a structured alignment, simple contrastive alignment should suffice. However, replacing our dual-level OT with CLIP-style transformation yields an inferior accuracy of 0.677 eV, suggesting that contrastive objectives alone are insufficient for cross-modal regression tasks, as they do not enforce a structured transformation that preserves geometric relationships. Furthermore, removing group-level transformation increases the MAE to 0.703 eV, while ablating instance-level transformation yields an MAE of 0.692 eV. Both are significantly worse than the full model, confirming that neither level alone can achieve the complete knowledge distillation required for accurate energy prediction. These results validate the superiority of our transport-based design in explicitly aligning the distributions and establishing fine-grained cross-modal correspondences.

To further investigate how the transport mechanisms contribute to cross-modal structural alignment, we quantify the structural discrepancy between the transported textual representations and their geometric counterparts. Using the notation introduced in the Methodology, we define the relative structural distortion error as $\epsilon_{\mathrm{rel}} = \epsilon / \|\mathbf{G}_{\mathrm{agg}}^\top \mathbf{G}_{\mathrm{agg}}\|_F$, where $\epsilon = \|\mathbf{G}_{\mathrm{agg}}^\top \mathbf{G}_{\mathrm{agg}} - \mathbf{U}_{\mathrm{agg}}^\top \mathbf{U}_{\mathrm{agg}}\|_F$ measures the discrepancy between the aggregated geometric and transported textual features. Here, $\mathbf{U}_{\mathrm{agg}} = [\mathbf{U}_1\mathbf{Q}_{11}, \mathbf{U}_2\mathbf{Q}_{22}, \ldots, \mathbf{U}_S\mathbf{Q}_{SS}]$ and $\mathbf{G}_{\mathrm{agg}} = [\mathbf{G}_1, \mathbf{G}_2, \ldots, \mathbf{G}_S]$. This metric measures the extent to which the pairwise inner-product structures of the two modalities are aligned. We compare the relative structural distortion across \emph{Taco} and its ablated variants to examine the respective contributions of the two transport levels to cross-modal structural alignment.
As reported in Figure \ref{tacoTSNE}(a), the full Taco model achieves a relative structural distortion error of 7.35\%. Removing the group-level transformation increases this error to 11.85\%, while removing the instance-level transformation raises it further to 14.39\%.


Our proposed GCOT transforms cross-modal features in a group-wise manner. This is motivated by the t-SNE visualization in Figure~\ref{tacoTSNE}(b), where the geometric features from OC data roughly aggregate into 9 clusters. These clusters are not clearly separated; therefore, we design a soft grouping strategy in this work rather than hard assignment. The performance of our method inherently depends on the choice of the group number $S$. To investigate this dependency, we conduct a sensitivity analysis by varying the number of groups from 1 to 15, with the results reported in Figure~\ref{tacoAblation}(c). The experimental findings show that using only one group leads to the highest MAE, indicating insufficient capacity for cross-modal transformation. When the group number exceeds 7, the MAE no longer decreases significantly, while the computational cost of optimal transport grows considerably. Based on this trade‑off and visualization
in Figure~\ref{tacoTSNE}(b), we set the group number $S$ to 9.

Finally, to validate the effectiveness of the proposed solver for the transport cost optimization, we plot its convergence trajectory. We track the difference between the transformation matrices obtained at consecutive iterations, defined as $\tau=\| \mathbf{R}^{(t)} - \mathbf{R}^{(t-1)} \|_F$, where $\mathbf{R}^{(t)}$ denotes the transformation at iteration $t$. The solver is considered converged when this difference falls below a sufficiently small $\tau$ and no further significant change is observed. As shown in Figure~\ref{tacoTSNE}(c), the proposed solver converges rapidly, and the difference between consecutive iterations diminishes to approximately $10^{-16}$, indicating a highly stable solution.

\section{Methodology}
\subsection{Encoders}
\textbf{Geometric encoder.}
As adsorption energy is an invariant scalar, we design our model to operate entirely within an invariant framework. High-order geometric descriptors are constructed to capture local environments, and all message-passing updates are performed on scalar representations.
Specifically, we construct a hierarchy of invariant geometric descriptors from local atomic environments.
The local environment of each atom is represented by descriptors of increasing representation order. The two-order descriptor captures radial interactions,
\begin{equation}
\mathbf g_i^{(2)}
=
\frac{1}{|\mathcal N(i)|}
\sum_{j\in\mathcal N(i)}
\phi_2(d_{ij}),
\end{equation}
where $d_{ij}=\|\mathbf r_{ij}\|$, and $\phi_2$ consists of a radial basis expansion followed by a multilayer perceptron.
To encode angular geometry, we further construct a three-order descriptor,
\begin{equation}
\mathbf g_i^{(3)}=\frac{1}{|\mathcal A(i)|}\sum_{(j,k)\in\mathcal A(i)}\phi_3\left(d_{ij},d_{ik},\cos\theta_{jik}\right),
\end{equation}
where $\mathcal A(i)=\left\{(j,k)\mid j,k\in\mathcal N(i),~j\neq k \right\}$ and $\cos\theta_{jik}=\frac{\mathbf r_{ij}^{\top}\mathbf r_{ik}}{\|\mathbf r_{ij}\|\,\|\mathbf r_{ik}\|}.$
Rather than explicitly enumerating increasingly expensive higher-order tuples, descriptors of high order $s\ge4$ are recursively constructed by composing lower-order invariant descriptors,
\begin{equation}
\mathbf g_i^{(s)}=\phi_s\left(\mathbf g_i^{(2)},\mathbf g_i^{(3)},\ldots,\mathbf g_i^{(s-1)}\right),\qquad s\ge4,
\end{equation}
where $\phi_s$ denotes a shared multilayer perceptron.
This hierarchical composition progressively captures increasingly complex coordination motifs while avoiding the exponential complexity associated with explicit higher-order enumeration.
The complete local geometric representation is obtained by
\begin{equation}
\mathbf h_i^{(0)}=\phi_f\left(\bigoplus_{s=2}^{S=5}\mathbf g_i^{(s)}\right),
\end{equation}
where $\oplus$ denotes feature concatenation and $\phi_f$ projects descriptors from different orders into a common latent space.

The hierarchical descriptors are subsequently refined through scalar message passing,
\begin{equation}
\mathbf m_{ij}^{(t)}=\phi_m\left(\mathbf h_i^{(t)}\oplus \mathbf h_j^{(t)}\oplus\phi_e(d_{ij})\right),
\end{equation}
where $\phi_e$ encodes pairwise distances using radial basis functions.
Node representations are updated by Layer Normalization (LN)
\begin{equation}
\mathbf h_i^{(t+1)}=\mathrm{LN}\left(\mathbf h_i^{(t)}+\sum_{j\in\mathcal N(i)}\mathbf m_{ij}^{(t)}\right).
\end{equation}

Since all geometric descriptors are constructed exclusively from invariant quantities and message passing involves only scalar features, the learned representations remain invariant.
After $T$ message-passing layers, the graph representation is obtained by mean pooling, $\mathbf g=\frac1{|\mathcal V|}\sum_{i\in\mathcal V}\mathbf h_i^{(T)}$.

\textbf{Text encoder.} 
The textual descriptors are processed by a text encoder based on the RoBERTa architecture~\cite{delobelle2020robbert}. 
Given initial token embeddings $\mathbf{u}^{(0)}=\mathcal{S}$, we use a stack of $M$ Transformer layers containing multi-head self-attention (MHA) to encode the textual representations. 
The update rule for the textual tokens at layer $m$ is: $\mathbf{u}^{(m+1)} = \text{MLP}(\text{MHA}(\mathbf{u}^{(m)}))$, where $\text{MHA}$ captures long-range dependencies among the chemical descriptors, and $\text{MLP}$ is responsible for the non-linear transformations of token features. 

\subsection{Group-Conditioned Optimal Transport}

\textbf{Soft group assignment.}
Direct instance-level alignment does not explicitly account for the potentially different structural organizations of textual and geometric representations. Since the two modalities are independently encoded, their samples may exhibit distinct distributional patterns even when individual instances are paired. We therefore introduce modality-specific soft grouping to characterize the latent organization of each representation space before establishing cross-modal correspondences. Rather than imposing rigid partitions, we assign each sample to multiple groups with different membership degrees, allowing the resulting group distributions to capture overlapping representation patterns.

Specifically, we compute membership scores that indicate the degree to which each sample belongs to each group. This soft assignment preserves uncertainty for samples near group boundaries, thereby enhancing transformation stability. Given the full textual feature matrix $\bar{\mathbf{U}}=[\mathbf{u}_1,\dots,\mathbf{u}_{N}]\in\mathbb{R}^{d\times N}$, we initialize learnable text group prototypes $\mathbf{C}^{\text{text}}=[\mathbf{c}_1^{\text{text}},\dots,\mathbf{c}_S^{\text{text}}]\in\mathbb{R}^{d\times S}$. Then, we define the textual soft assignment matrix $\mathbf{A}\in\mathbb{R}^{N\times S}$ with row-wise softmax normalization: 
\begin{equation}
\mathbf{A}_{ki}=\frac{\exp(\mathbf{u}_k^\top \mathbf{c}_i^{\text{text}})}
{\sum_{i'=1}^{S}\exp(\mathbf{u}_k^\top \mathbf{c}_{i'}^{\text{text}})},
\end{equation}
where $\mathbf{A}_{ki}$ represents the membership degree of textual sample $k$ belonging to the $i$-th text group. 
Larger entry values of $\mathbf{A}$ indicate higher similarity between the sample embedding and the corresponding group prototype. 

We optimize text prototypes by jointly minimizing a reconstruction loss and maximizing an entropy regularization term, which avoids trivial collapsed assignments where all samples fall into a single group:
\begin{equation}
\min_{\mathbf{C}^{\text{text}}}
\left\|
\mathbf{A} \left(\mathbf{C}^{\text{text}}\right)^\top-\bar{\mathbf{U}}^\top
\right\|_{F}^{2}
-
\lambda_h \frac{1}{N}\sum_{k=1}^{N} H(\mathbf{A}_{k,:}),
\label{eq:soft_assign_obj}
\end{equation}
where $\lambda_h$ controls entropy regularization strength, and $H(\cdot)$ stands for Shannon entropy. The reconstruction term drives group centers to capture inherent text representation patterns, while the entropy term preserves assignment diversity and facilitates stable group-level matching. The entropy regularization encourages non-deterministic assignments, allowing each sample to retain graded memberships across multiple groups. Together with the reconstruction objective, it provides a soft characterization of the representation space rather than enforcing hard sample-to-group assignments.

Similarly, we derive geometric soft assignment based on the geometric feature matrix $\bar{\mathbf{G}}=[\mathbf{g}_1,\dots,\mathbf{g}_{N}]\in\mathbb{R}^{d\times N}$ and geometric prototypes $\mathbf{C}^{\text{geo}}=[\mathbf{c}_1^{\text{geo}},\dots,\mathbf{c}_S^{\text{geo}}]\in\mathbb{R}^{d\times S}$. The geometric assignment matrix $\mathbf{B}\in\mathbb{R}^{N\times S}$ is formulated as:
\begin{equation}
\mathbf{B}_{lj}=\frac{\exp(\mathbf{g}_l^\top \mathbf{c}_j^{\text{geo}})}
{\sum_{j'=1}^{S}\exp(\mathbf{g}_l^\top \mathbf{c}_{j'}^{\text{geo}})},
\end{equation}
where $\mathbf{B}_{lj}$ quantifies the membership of geometric sample $l$ to the $j$-th geometric group.

\textbf{Soft group-based modality transformation.}
With soft assignment matrices $\mathbf{A}$ and $\mathbf{B}$ obtained, we construct grouped feature sets $\{\mathbf{U}_i \in \mathbb{R}^{d\times N}\}_{i=1}^S$ and $\{\mathbf{G}_j \in \mathbb{R}^{d\times N}\}_{j=1}^S$. 
Each $\mathbf{U}_i$ stacks the features of all textual samples as columns (each column is a $d$-dimensional embedding vector), and each $\mathbf{G}_j$ contains the full set of geometric samples arranged in the same column-wise fashion.
All groups share the total sample number $N$, as different groups are distinguished only by their soft membership weights, not by hard partitions.

To conduct group-wise distribution comparison using the Wasserstein distance from optimal transport, each group needs to be formulated as a probability distribution defined over the set of samples. We thus transform membership scores into probability distributions such that each group corresponds to a valid probability measure supported on all samples. Specifically, to convert membership scores into probability distributions, we first compute the group-level empirical probability measures $\mu_i$ and $\nu_j$ for textual groups and geometric groups, respectively.
The underlying intuition is to treat each group as a discrete distribution over sample points, where the probability mass assigned to each sample is proportional to its degree of membership to that group.
\begin{equation}
\mu_i := \sum_{k=1}^{N} \mathbf{a}^{(i)}_{k} \delta_{\mathbf{U}_i(k)}, \quad
\nu_j :=\sum_{l=1}^{N} \mathbf{b}_{l}^{(j)} \delta_{\mathbf{G}_j(l)},
\label{eq:cluster_measures}
\end{equation}
where $\mathbf{U}_i(k)$ and $\mathbf{G}_j(l)$ denote the $k$-th column of $\mathbf{U}_i$ and the $l$-th column of $\mathbf{G}_j$, i.e., the feature vectors of the $k$-th textual sample and the $l$-th geometric sample, respectively, and $\delta_{\mathbf{x}}$ denotes the Dirac measure concentrated at point $\mathbf{x}$. 
The normalized weights are obtained from the soft assignment matrices $\mathbf{A}$ and $\mathbf{B}$:
\begin{equation}
\mathbf{a}_k^{(i)}=\frac{\mathbf{A}_{ki}}{\sum_{k'} \mathbf{A}_{k'i}}, \quad\mathbf{b}_l^{(j)}=\frac{\mathbf{B}_{lj}}{\sum_{l'} \mathbf{B}_{l'j}}.
\end{equation}
The normalization ensures that $\sum_k \mathbf{a}_k^{(i)} = 1$ and $\sum_l \mathbf{b}_l^{(j)} = 1$, so that each $\mu_i$ and $\nu_j$ is a valid probability distribution. 
Intuitively, the weight $\mathbf{a}_k^{(i)}$ tells us how representative sample $k$ is for group $i$, and points with low membership contribute very little mass.

To compare a textual group $\mu_i$ with a geometric group $\nu_j$, we use the squared $2$-Wasserstein distance, which measures the minimal total squared Euclidean cost of transporting the mass of $\mu_i$ onto $\nu_j$:
\begin{equation}
W_2^2(\mu_i, \nu_j) := \min_{\mathbf{Q}\in\mathcal{U}(\mathbf{a}^{(i)},\mathbf{b}^{(j)})} \sum_{k=1}^{N} \sum_{l=1}^{N} \mathbf{Q}(k,l)\,\|\mathbf{U}_i(k)-\mathbf{G}_j(l)\|_2^2,
\label{eq:wasserstein}
\end{equation}
where the transport matrix $\mathbf{Q} \in \mathbb{R}^{N \times N}$ specifies how much mass moves from textual point $k$ to geometric point $l$.
The admissible transport plans must respect the marginal masses exactly:
\[
\sum_{l} \mathbf{Q}(k,l) = \mathbf{a}_k^{(i)},\quad
\sum_{k} \mathbf{Q}(k,l) = \mathbf{b}_l^{(j)},\quad
\mathbf{Q} \ge \mathbf{0}.
\]
We formalize this set, the transport polytope, as
\begin{equation}
\mathcal{U}(\mathbf{a},\mathbf{b}) := \{\mathbf{Q}\in\mathbb{R}_+^{N \times N}: \mathbf{Q}\mathbf{1}_N = \mathbf{a},\ \mathbf{Q}^\top\mathbf{1}_N = \mathbf{b}\},
\label{eq:transport_polytope}
\end{equation}
with the all-one vector $\mathbf{1}_N$ of length $N$. 

It is worth noting that, for each pair of text and geometry groups \((i,j)\), we allow the instance-level transport plan \(Q_{ij}\) to establish correspondences between different samples across the two modalities, rather than restricting it to the original paired samples. Specifically, an entry \(Q_{ij}^{kl}\) can be nonzero even when \(k\neq l\), allowing the \(k\)-th textual representation to transport probability mass to the \(l\)-th geometric representation. This design is motivated by two considerations. First, although the textual and geometric representations are derived from paired configurations, their independently learned encoders may induce different local neighborhood structures in the respective representation spaces. Consequently, enforcing the original one-to-one sample correspondences may not yield an appropriate alignment between the two latent distributions. Second, the group-conditioned distributions \(a^{(i)}\) and \(b^{(j)}\) may have different membership weights. In this case, restricting \(Q_{ij}\) to a diagonal coupling would require the two marginal distributions to coincide, which is not generally guaranteed. Allowing non-diagonal transport therefore provides the flexibility to accommodate modality-specific representation structures while satisfying the prescribed marginal constraints. Importantly, this cross-sample transport does not change the identities of the original configurations; it captures their distributional correspondences in the latent spaces rather than imposing strict instance-level equivalence.

We first transform groups at a coarse level, then refine within each group at the instance level. We therefore formulate a Wasserstein method that jointly learns a transformation $T$ and the group-correspondence matrix $\mathbf{P}$:
\begin{equation}
\min_{\mathbf{P}\in\mathcal{B}_S,\;T\in\mathcal{T}} \sum_{i=1}^S \sum_{j=1}^S \mathbf{P}_{ij}\,W_2^2\bigl(T(\mu_i), \nu_j\bigr),
    \label{eq:hierarchical_ot}
\end{equation}
where $\mathbf{P} \in \mathbb{R}^{S \times S}$ encodes the strength of correspondences between soft groups.
A large $\mathbf{P}_{ij}$ indicates a strong link between textual group $i$ and geometric group $j$; a small value indicates a weak or absent link.
The set $\mathcal{B}_S := \mathcal{U}(\mathbf{1}_S/S, \mathbf{1}_S/S)$ is the $S$-th Birkhoff polytope (doubly stochastic matrices).

This formulation establishes a hierarchical correspondence between the two modalities. At the outer level, the doubly stochastic matrix $\mathbf{P}$ captures the distributional relationships between textual and geometric groups, accounting for the potentially different organizations of their respective representation spaces. Conditioned on each group pair, the inner Wasserstein distance further determines how probability mass should be transported between individual samples. In this way, the outer optimization captures cross-modal structural correspondence, while the inner optimization resolves fine-grained instance-level distributional matching.
The two levels are complementary. Group-level matching alone cannot resolve the detailed correspondence between the underlying sample distributions, whereas instance-level matching without group conditioning does not explicitly account for the global organization of the two representation spaces. Their joint optimization thus provides a structured distributional transformation that considers both cross-modal organization and sample-level correspondence.

\subsection{Optimization for Transport Costs}

\textbf{Optimization problem formulation.} 
Recall that in the formulation~(\ref{eq:hierarchical_ot}), we need to learn a transformation $T$ that maps each textual group distribution into the geometric feature space.
We now specify the form of $T$: A natural choice is a linear transformation. However, the transformation should establish cross-modal correspondence without unnecessarily deforming the intrinsic geometry of the textual representation space. An unconstrained linear mapping may arbitrarily alter pairwise distances and angles among textual embeddings, introducing additional structural distortion during distributional matching. The Stiefel constraint preserves Euclidean geometry. We therefore restrict \(T\) to the Stiefel manifold \(\mathcal{V}_{d,d}\), the set of all \(d\times d\) orthogonal matrices:
\begin{equation}
\mathcal{V}_{d,d} := \{\mathbf{R}\in\mathbb{R}^{d\times d}: \mathbf{R}^\top \mathbf{R} = \mathbf{I}\}.
\label{eq:stiefel_def}
\end{equation}
For any $\mathbf{R} \in \mathcal{V}_{d,d}$, applying $\mathbf{R}$ to a textual feature vector simply rotates or reflects it.
This Stiefel constraint allows rotations and reflections while preserving pairwise Euclidean distances and inner products within the source representation space. It thus provides a geometry-preserving inductive bias for cross-modal representation transformation.

With this choice, the general transformation $T(\mu_i)$ becomes the transformed distribution obtained by replacing each support point $\mathbf{U}_i(k)$ with $\mathbf{R} \mathbf{U}_i(k)$, while keeping the same probability weights $\mathbf{a}^{(i)}$.
Plugging this into the problem~(\ref{eq:hierarchical_ot}) yields the concrete optimization:
\begin{equation}
    \min_{\mathbf{P},\mathbf{R},\{\mathbf{Q}_{ij}\}} \sum_{i,j} \mathbf{P}_{ij} C_{ij}(\mathbf{R},\mathbf{Q}_{ij}) 
    \quad \text{s.t.} \quad 
    \begin{aligned}
    &\mathbf{P}\in\mathcal{B}_S,\; \mathbf{R}\in\mathcal{V}_{d,d},\; \\
    &\mathbf{Q}_{ij}\in\mathcal{U}(\mathbf{a}^{(i)},\mathbf{b}^{(j)}),
    \end{aligned}
    \label{eq:stiefel_align}
\end{equation}
with the pairwise group cost $C_{ij}(\mathbf{R},\mathbf{Q}_{ij})$ defined as
\begin{equation}
    C_{ij}(\mathbf{R},\mathbf{Q}_{ij}) := \sum_{k,l} \mathbf{Q}_{ij}(k,l) \|\mathbf{R}\mathbf{U}_i(k)-\mathbf{G}_j(l)\|_2^2,
    \label{eq:cost}
\end{equation}
where $C_{ij}$ measures the squared $2$-Wasserstein distance between the rotated textual group $i$ and the geometric group $j$, using the transport plan $\mathbf{Q}_{ij}$.
As $\mathbf{R}$ is orthogonal, the norm of any transformed vector remains unchanged, so the intrinsic structure of the textual features is preserved during transformation.

Directly solving formulation (\ref{eq:stiefel_align}) is computationally demanding because the Wasserstein distances involve constrained optimization over $\mathbf{Q}_{ij}$, and the entire problem is non‑convex.
To overcome this, we introduce entropic regularization on all transport couplings.
This replaces the exact Wasserstein distances with their Sinkhorn counterparts, which can be computed efficiently via iterative matrix scaling and lead to smoother optimization landscapes with better sample complexity.
Omitting the constraints for brevity, the final objective becomes
\begin{equation}
    \min_{\mathbf{P},\mathbf{R},\{\mathbf{Q}_{ij}\}} \sum_{i,j} \left( \mathbf{P}_{ij} C_{ij}(\mathbf{R},\mathbf{Q}_{ij}) -  H(\mathbf{Q}_{ij}) \right) -  H(\mathbf{P}), 
    \label{eq:final_objective}
\end{equation}
where $H(\cdot)$ denotes the Shannon entropy.
The entropy terms penalize overly deterministic transport plans, encouraging the model to spread mass more uniformly.

\textbf{Optimization solver.} 
Problem (\ref{eq:final_objective}) is difficult to solve directly because it is non‑convex: the variables $\mathbf{P}$, $\mathbf{R}$, and $\{\mathbf{Q}_{ij}\}$ are multiplied together in the objective, and $\mathbf{R}$ is constrained to lie on the Stiefel manifold $\mathcal{V}_{d,d}$, which is not a convex set.
Although the alternating direction method of multipliers (ADMM) was originally developed as a convergent solver for convex problems~\cite{wang2019global,neal2011distributed}, it has been successfully applied to many non‑convex settings~\cite{eckstein1992douglas} where the problem structure allows efficient local updates.
A key structural feature of our problem is that the objective separates naturally across group pairs $(i,j)$: each term $\mathbf{P}_{ij} C_{ij}(\mathbf{R},\mathbf{Q}_{ij}) - H(\mathbf{Q}_{ij})$ involves only the local coupling $\mathbf{Q}_{ij}$ and the global rotation $\mathbf{R}$, with no direct coupling between different $(i,j)$ blocks besides the shared $\mathbf{R}$.
This immediately suggests a divide‑and‑conquer strategy: we can assign each group pair its own local rotation estimate $\mathbf{R}_{ij}$, allow them to solve for an optimal $\mathbf{R}_{ij}$ independently, and then enforce consensus among all local estimates to obtain a single global rotation $\tilde{\mathbf{R}}$.

To implement this idea, we introduce auxiliary variables $\mathbf{R}_{ij}$ for each group pair and explicitly require them to agree with a common global variable $\tilde{\mathbf{R}}$.
The problem then becomes
\begin{equation}
    \begin{aligned}
\min_{\substack{\mathbf{P},\tilde{\mathbf{R}}\\ \{\mathbf{R}_{ij},\mathbf{Q}_{ij}\}}} \sum_{i,j} \Bigl( 
    &\mathbf{P}_{ij} C_{ij}(\mathbf{R}_{ij},\mathbf{Q}_{ij}) - H(\mathbf{Q}_{ij}) \Bigr) -  H(\mathbf{P}) \\
    &\text{s.t.} \quad \mathbf{R}_{ij} = \tilde{\mathbf{R}},\;\forall i,j,
    \end{aligned}
    \label{eq:split_admm}
\end{equation}
where the set constraints on $\mathbf{P}$, $\mathbf{R}_{ij}$, and $\mathbf{Q}_{ij}$ are omitted for brevity.
The equality constraints $\mathbf{R}_{ij} = \tilde{\mathbf{R}}$ are the glue that forces all local rotations to ultimately agree.
We then form the scaled augmented Lagrangian to handle these constraints in a numerically stable fashion:
\begin{equation}
\begin{aligned}
\mathcal{L}_{map} = \sum_{i,j} \Bigl( 
& \mathbf{P}_{ij} C_{ij}(\mathbf{R}_{ij},\mathbf{Q}_{ij}) 
+ \frac{\mu}{2}\|\mathbf{R}_{ij}-\tilde{\mathbf{R}} \|_F^2 \\
& + \langle \mu\mathbf{\Lambda}_{ij},\mathbf{R}_{ij}-\mathbf{R} \rangle 
-  H(\mathbf{Q}_{ij}) \Bigr) 
-  H(\mathbf{P}),
\end{aligned}
\label{eq:admm_lagrangian}
\end{equation}
where $\mu>0$ is the ADMM penalty parameter that controls the strength of consensus enforcement, and $\{\mathbf{\Lambda}_{ij}\}$ are the scaled Lagrange multiplier matrices associated with each constraint $\mathbf{R}_{ij} = \tilde{\mathbf{R}}$. Full details of the ADMM algorithm are presented in Section \ref{sec:admm_details} of the Supplementary Materials.

\subsection{Learning Objectives}

The group‑conditioned optimal transport module not only transforms the distributions of the two modalities, but also produces a concrete mapping that transports textual features into the geometric latent space.
This mapping, i.e., the text-geometry projector, is given by the learned transformation $\mathbf{R} \in \mathcal{V}_{d,d}$, which we obtained by solving problem~(\ref{eq:final_objective}).
Because $\mathbf{R}$ is orthogonal, it preserves norms and inner products; applying it to a textual embedding rotates the vector into the geometric coordinate system without distorting its structure.
For any textual sample $i$ with text embedding $\mathbf{u}_i$, the transformed geometric‑aware representation is simply $\hat{\mathbf{g}}_i = \mathbf{R}\,\mathbf{u}_i$.
During the training stage, this transformation is optimized; during inference, it remains fixed.

With the transformed embeddings $\hat{\mathbf{g}}_i$ in hand, the adsorption energy prediction reduces to a standard regression problem.
An MLP prediction head $h(\cdot)$ maps each $\hat{\mathbf{g}}_i$ to a scalar energy.
To make training stable, we first standardize the ground‑truth adsorption energies using z‑score normalization: $\bar{y}_i = (y_i - \mu)/\sigma$,
where $\mu$ and $\sigma$ are the mean and standard deviation computed over the entire training set.
The head therefore predicts normalized values $\hat{\bar{y}}_i = h(\hat{\mathbf{g}}_i)$, and we train it with the mean absolute error loss, which is robust to outliers and directly measures average prediction accuracy in the normalized space:
$\mathcal{L}_{\mathrm{energy}} = \frac{1}{|\mathcal{B}|}\sum_{i\in\mathcal{B}}\bigl|\hat{\bar{y}}_i - \bar{y}_i\bigr|$, where $\mathcal{B}$ denotes the current mini‑batch.
The full training objective combines the transformation loss and the energy regression loss: $\mathcal{L} = \mathcal{L}_{\mathrm{map}} + \lambda\,\mathcal{L}_{\mathrm{energy}}$, where $\mathcal{L}_{\mathrm{map}}$ is the optimal transport loss from the GCOT module.

\subsection{Textual Descriptor-Based Inference}

Once training is complete, the model can predict adsorption energies directly from textual input. For a textual descriptor with embedding $\mathbf{u}^*$, we apply the frozen Stiefel transformation $\mathbf{R}$ to project it into the geometric latent space as $\hat{\mathbf{g}}^* = \mathbf{R}\,\mathbf{u}^*$. The transformed feature is then fed into the energy prediction head to obtain the predicted energy. The entire pipeline therefore accepts a textual description and returns an adsorption energy, enabling efficient prediction that inherits the structural knowledge distilled from 3D configurations through GCOT transformation.

The above design offers two key advantages in terms of both efficiency and accuracy. 1) The optimal transport module in GCOT is employed only during training to establish the cross-modal transformation. At test time, the learned projection matrix $\mathbf{R}$ is frozen, and the inference pipeline reduces to just three components: a text encoder, the orthogonal projection $\mathbf{R}$, and an MLP energy head. The forward pass operates exclusively on textual descriptors and bypasses the computationally expensive iterative message-passing of the geometric encoder. It should be noted that our approach does not attempt to circumvent the use of 3D structural information, as the textual descriptors are derived from 3D data; rather, it distills geometric knowledge into the text domain during training, enabling efficient and accurate predictions at inference time without invoking the geometric encoder. 2) The learned transformation $\mathbf{R}$ is a linear orthogonal map that rotates text embeddings without distorting their semantic structure. This rotation aligns the text features with the energy-sensitive directions that the geometric encoder has learned from 3D data, enabling the model to maintain high accuracy using only textual descriptors.

This method also has certain limitations. Since inference relies solely on textual descriptors, \emph{Taco} inherits the intrinsic information bottleneck of descriptor-based representations. Specifically, adsorption configurations that share the same textual descriptor (e.g., identical adsorption sites but different molecular orientations) cannot be distinguished at inference time. Nevertheless, this limitation is alleviated in our studied IS2RE setting. These initial configurations within the same adsorption site but different molecular orientations typically converge to the same relaxed configurations and therefore lead to negligible differences in adsorption energy. 
Specifically, in OC20 IS2RE, only two configurations exhibited a descriptor collision, where different geometries were mapped to the same textual descriptor; the energy difference between these two configurations was 0.00095 eV. Moreover, no descriptor collisions were identified in OC20Dense. These results indicate that descriptor collisions are extremely rare, suggesting that the inherent limitation of descriptor-based inference is substantially alleviated.

\section{Conclusion}
In this paper, we presented \emph{Taco}, a cross-modal framework that learns a text-to-geometry representation transformation during training, enabling adsorption energy prediction from compact textual inputs at inference time. We formulated this cross-modal transformation as an optimization problem, seeking a projection that preserves the intrinsic structure of both modalities. Based on this formulation, we introduced a group-conditioned optimal transport module that operates in two stages: it first transforms the two modalities at the group level, then refines the correspondence at the instance level. This coarse-to-fine strategy bridges the textual-structural gap and transfers geometric knowledge into textual representations, allowing accurate prediction using text inputs.
Extensive experiments on the Open Catalyst benchmark have demonstrated that \emph{Taco} achieves competitive prediction performance compared with the state-of-the-art geometry-based methods, while being more efficient than the geometry-based counterparts.

\section{Code and Reproducibility}

The source code will be made publicly available in a public repository upon acceptance of this manuscript.


\bibliography{references}

\end{document}